# COMPARISON OF CERTIFICATE POLICIES FOR MERGING PUBLIC KEY INFRASTRUCTURESDURING MERGER AND ACQUISITION OF COMPANIES

Balachandra Muniyal[1], Prema K.V[2], Mamatha Balachandra[3]

[1]Dept. of Information and Communication Technology, Manipal Institute of Technology, Manipal University, Manipal,India
bala_muniyal@yahoo.com
[2]Dept. of Computer Science and Engineering, Modi Institute of Technological Sciences, Jaipur, India
drprema.mits@gmail.com
[3]Dept. of Computer Science and Engineering, Manipal Institute of Technology, Manipal University, Manipal,India
mamthabc@yahoo.co.in

## ABSTRACT

*The Public Key Infrastructure(PKI) provides facilities for data encryption, digital signature and time stamping. It is a system where different authorities verify and authenticate the validity of each participant with the use of digital certificates. A Certificate Policy (CP) is a named set of rules and it indicates the applicability of a certificate in a Public Key Infrastructure. Sometimes two companies or organizations with different PKIs merge. Therefore it would be necessary that their PKIs are also able to merge. Sometimes, the unification of different PKIs is not possible because of the different certificate policies. This paper presents a method to compare and assess certificate policies during merger and acquisition of companies.*

## 1. INTRODUCTION

In order to transmit critical data safely over the internet during e-commerce and e-business transactions, robust and trustworthy security systems are required. Public Key Infrastructure(PKI) [1][2][3]is a security infrastructure that provides the necessary security services in enterprises. One of the main goals of PKI is the verification andauthenticationof each participant in the business with the use of digital certificates. A Certificate Policy(CP) is a named set of rules that indicates the applicability of a certificate to a particular community and/or class of application with common security requirements. An X.509 Version 3 certificate may identify a specific applicable CP, which may be used by a relying party to decide whether or not to trust a certificate, associated public key, or any digital signature verified using the public key for a particular purpose. A Certificate Practice Statement (CPS)states how a certificate authority im-plements a CP[4]. They have a commonoutline, as indicated in the IETF Request for Comments(RFC)3647 [5] or in the older version, RFC2527 [6]. The policy indicates the PKI certificates' profile and the architecturalstructure of the underlying trusted third party. Thus a certificate authority issues acertificate to an end user with a specific certificate policy[7]. If different end userswith different domains want to establish a secure communication, they need to find away to trust each other. That means, their certificates need to follow thesame standard. X.509 is a standard that specifies the standard format for the certificates, the certificate revocation lists and the certificate path validation algorithm.

When users from different domains want to communicate, interoperability between PKIs is a major issue to be considered. Interoperability between PKIs makes possible the secureinterconnection and co-operation between different PKI structures.PKI interoperability is usually addressed through the cross-certification service, whichcan be described as the way to establish chains of trust between different certificationauthorities (CAs)[8][9][10]. However, cross-certification is not yet technically providedin an automated way, resulting often in a difficult and time-consuming paper-basedprocess, which reduces the flexibility and usability of the method itself. The lackof automated cross-certification is largely due to the inadequate standardization ofthe certificate policies (CP), which define the PKI certificates' profile and, thus, formthe basic comparison criteria for the mutual acceptance of CAs[6]. More specifically,although the CP structure is defined in some of the existing standards[11-16], there is stilla significant gap in the standardization of the CP content (e.g., roles of the involvedsubjects, certification and registration requirements etc.). In addition, there is nosystemized way for the development and the comparison of CPs, thus making theircomparative analysis a difficult task. The above restrictions together with the lackof the necessary legal/regulatory PKI harmonization[17], disable the automated CPcomparison, obstructing in this way the automation of the overall cross-certificationservice and, thus, making difficult the secure electronic co-operation, information ex-change and knowledge sharing.

The trust depends upon the content of their certificates. In general, the certificatepolicy is optional and is indicated as an extension fieldin the X.509 certificate. The policy extension field in the certificate indicates the policywith an object identifier (OID).

## 1.1 Unification of Public Key Infrastructures

When two or more business corporations are collaborated or during their acquisitions, merging of their PKIs at the root is a simple and straight forward approach[18][19]. This is the best solution when the interoperability between the PKIs is temporary and dynamically change with the market requirements. The merging process needs to be low-cost, easily constructed and flexible. Figure 1 shows, the unification of two Hierarchical PKIs through cross certification.

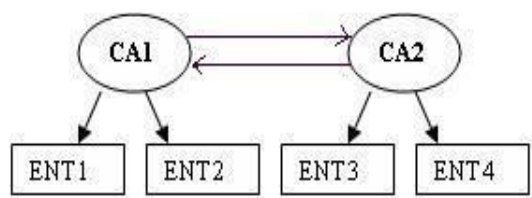

Figure 1. Merged Hierarchical PKIs with cross-certifications

In order to merge two infrastructures, the certificate policies should be merged as well. If the certificate policies are different, the unification of PKIs is not allowed.

This paper is organizedin the following way: Section 2 provides a brief overview of the PKI components. Section 3 highlights PKI models. Section 4 explains certificate policies, policy structure and their comparison by parsing process. Section 5 describes the formalization of the proposed method of certificate policy comparison and assessment to merge PKIs. Finally, section 6 concludes the paper.

## 2. PKI Components

PKI consists of a number of components like certification authorities, registration authorities, repositories and users. The users of the PKI can be divided into different categories such as certificate holders and relying parties.

- **Certification authority (CA):** The CA is the basic building block of the PKI. The CA confirms the identities of parties sending and receiving electronic payments or other communications. It is a collection of computer hardware, software and the people who operate it. The CA is known by two attributes[20]: its name and its public key. The CA performs 4 basic PKI functions: 1) Issues certificates(i.e. creates and signs them); 2)Maintains certificate status information and issues Certificate Revocation Lists(CRLs). 3)Publishes its current(e.g. unexpired) certificates and CRLs, so users can obtain the information they need to implement security services and 4) Maintains archives of status information about the expired certificates that it issues.

- **Registration authority (RA):** Once authentication is done by a CA it will ask the RA to register or vouch for the identity of users to a CA. The certificate contents are made in such a way that it will reflect the information presented by the requesting entity and sometimes they also reflect third party information. CA and RA are similar in a way that both contain computer hardware, software and an operator. But a small difference we can say that CA will be mostly operated by multi-user whereas RA will be often operated by a single user. Each CA contains a list of its trustworthy RAs. CA identifies RA by a name and a public key. RAs signature on a message means that a CA which has a trustworthy relation with that RA can trust the message. So the RA should be providing an adequate protection for its own private key.

- **PKI Repository:** It is a database for a CA where the digital certificates have been stored. When the users want to confirm the status of the digital certificate for any of the other reason they will contact the repository and the repository will in turn produces digitally signed messages and will send back to the user.

- **PKI Users:** PKI users are organizations or individuals that use the PKI. They rely on the other components of the PKI to obtain certificates, and to verify the certificates of other entities with whom they do business. End-entities include the *relying party*, who relies on the certificate to know with certainty the public key of another entity and the *certificate holder*, that issued a certificate and can sign digital documents. An individual or organization may be both a relying party and a certificate holder for various applications.

## 3. PKI Models

PKI trust models are also referred to as PKI structures or PKI architectures. Different business corporations deploy different types of PKIs. They are: Single CA, Hierarchical, Multirooted Hierarchical, Mesh , Bridgeand Hybridmodels.

## 3.1 Single CA PKI model

As shown in Figure 2, A Single CA PKI architecture is one that contains a single CA and provides the PKI services for all the users or the end entities (ENT1, ENT2, ENT3 and ENT4 in Figure 2) of the PKI. There is a single trust anchor that has to be trusted by all the users.

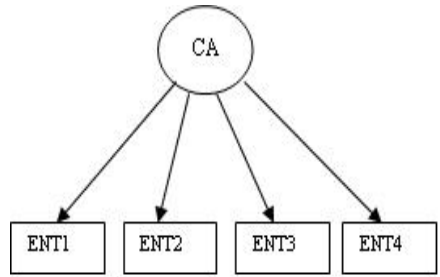

Figure 2. Single CA PKI architecture

If the CA's private key is compromised, the security of the entire system collapses. Even though this configuration is very easy to deploy, scalability is very poor if the community of users is very large.

## 3.2 Hierarchical PKI model

A Hierarchical PKI, as depicted in Figure 3, is one in which all of the subscribers / relying parties trust a single CA. This CA is called Root CA (RCA in Figure 3) and is the most trusted anchor. The Root CA certifies the public keys of subordinate CAs. These CAs (CA1 and CA2 in Figure 3) certify their subscribers or may, in a large PKI, certify other CAs. In this architecture, certificates are issued in only one direction, and a CA never certifies another CA "superior" to itself. Typically, only one superior CA certifies each CA. Certificate path construction in a Hierarchical PKI is a straightforward process that simply requires the relying party to successively retrieve issuer certificates until a certificate is located that was issued by the trusted root. Hierarchical PKIs are scalable; certification paths are easy to develop and certification paths are relatively short [23]. However, reliance on a single trust point may result in compromise of the entire PKI.

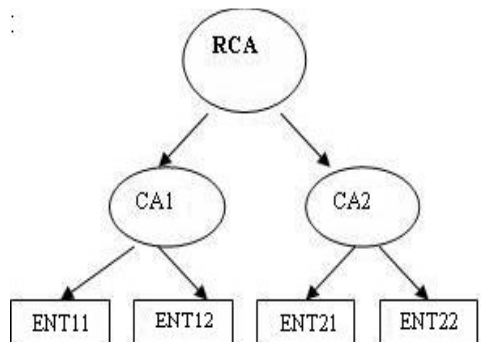

Figure 3. Hierarchical PKI

### 3.3 Multi-Rooted Hierarchical PKI

The multi-rooted Hierarchical PKI is used in popular web browsers. It is the widely used varia-tion on the single-rooted Hierarchical PKI. As shown in Figure 4, a multi-rooted Hierarchical PKI is formed by inclusion of multiple root CAs as a trust list. As far as certificate validation is concerned, there is no much difference between single-rooted Hierarchical PKI and multi-rooted Hierarchical PKI.  The difference is that a certificate will be accepted only if it can be verified back to any of the set of trust anchors in the trust list.

This scheme has some drawbacks as well. This approach is suitable only for applications wherein the number of certificate verifications is limited, otherwise it may introduce certain security vulnerabilities[25].  Users should have proper idea of the certificate policies and oper-ating practices of the various trust anchors. Also they must be aware of which root was used to verify a given certificate.  Additionally, the compromise of any trusted CA private  key or the insertion of an unwanted CA certificate to the trust list may  compromise the entire system. This can be an efficient solution for certificate path verification only if the trust list is properly managed and kept to a reasonable size.

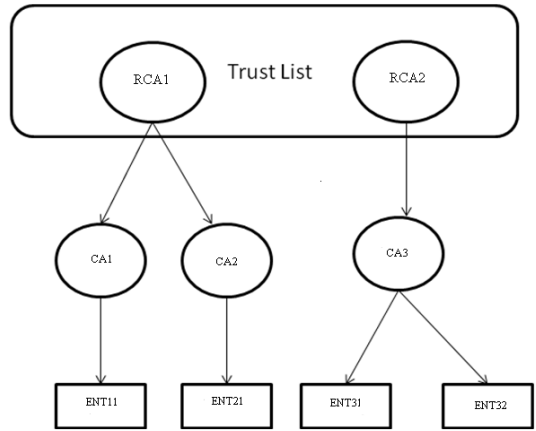

Figure 4. A multi-Rooted Hierarchical PKI

### 3.4 Mesh PKI model

A PKI constructed with peer-to-peer CA relationships is called a Mesh PKI or a "web of trust"[5]. In a mesh style PKI, as depicted in Figure 5, each subscriber trusts its own CA. The CAs in this environment have no superior/ subordinate relationship. In a mesh, CAs in the PKI cross certify each other. Figure 5 depicts a mesh PKI that is fully cross-certified, however, it is possible to construct and deploy a mesh PKI with a mixture of unidirectional and cross-certifications[26]. Compromise of a single CA cannot bring down the entire PKI. Mesh PKIs can easily incorporate a community of users. However, certification path construction in a mesh PKI is more complex than in a Hierarchical PKI due to the likely existence of multiple paths between a relying party's trust anchor and the certificate to be verified, and the potential for loops and cycles in non-hierarchical certificate graphs.

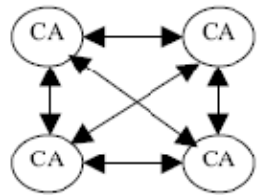

Figure 5. A Mesh PKI

## 3.5 Hybrid PKI model

Hybrid PKI is theinterconnection of different PKIs via cross certification. This enables relying parties of each to verify and accept certificates issued by the other PKI[27]. If the interconnection is between only Hierarchical PKIs, Root CAs of all the participating PKIs cross-certify each other facilitating interoperability between PKIs. Similarly, if the PKIs are mesh style, then a CA within each PKI is selected, more or less arbitrarily, to establish the cross-certification. This results in the creation of a larger mesh PKI. However, the participating PKIs need not be of the same type. Figure 6 depicts a hybrid situation resulting from a Hierarchical PKI cross-certifying a mesh PKI.

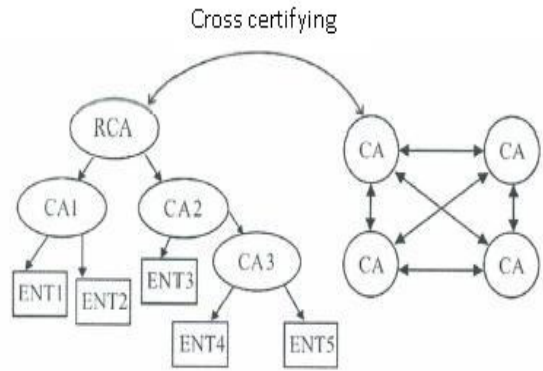

Figure 6. A Hybrid PKI

As the number of cross certified PKIs grows, the number of relationships between them grows exponentially resulting in complex certificate path verification.

## 3.6 Bridge PKI model

Another approach to the interconnection of PKIs is the use of a "bridge" certification authority (BCA). A BCA architecture was designed to address the shortcoming of Hierarchical and Mesh PKIs [28]. A BCA connects multiple PKIs to establish trust paths among them. The BCA is not intended to be used as a trust point by the users of the PKI. As shown in the Figure 7, the BCA cross-certifies with one CA (known as a "principal" CA [PCA]) in each participating PKI.

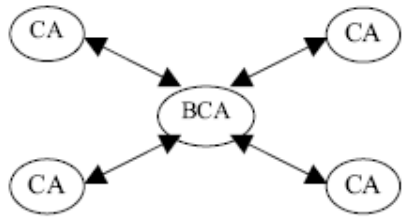

Figure 7. Bridge PKI

Since each PKI only cross-certifies with one other entity (i.e., the BCA), and the BCA cross-certifies only once with each participating PKI, the number of relationships in this environment grows linearly with the number of PKIs resulting in certification path discovery easier than mesh PKI. However, the certification path discovery in Bridge PKI model is more difficult than Hierarchical PKI. The sole purpose of BCA is to establish trust relationships between different PKIs. It is not considered as a trust point, rather it is considered as a trust arbitrator. The new CAs can be added just by a pair of certificates between BCA and the new CA itself.

In general, none of these architectures is perfect for all situations. Single CA PKI has the draw-back of relying on a single CA. In a Hierarchical PKI model, certificate path is unidirectional, so certificate path development and validation is simple and straight forward. However, if the root CA is compromised, which is everyone's trust point, the security of the whole system col-lapses. Mesh architecture is also widely used in applications such as MANET, but certificate path development is more complex than in a hierarchy. Unlike a hierarchy, building a certificate path from a user's certificate to a trust point is nondeterministic. The Bridge Certification Au-thority (BCA) architecture was designed to address the shortcomings of the Hierarchical and Mesh PKI architectures, and to link PKIs that implement different architectures, but certificate path discovery is not simple. Since Hybrid PKI is the mixture of different PKI architectures, the complexity of certificate path verification increases.

Therefore, the people who deploy PKI should choose the best architecture that is suited to their enterprise situation. A single CA is appropriate for a small community. A Hierarchical PKI is best suited to organizations with a well-defined structure. For organizations with no well-defined structure, a mesh PKI is the ideal one. Once the population increases, a bridge CA may be deployed but with extra effort to set up bridge CAs.

## 4. CERTIFICATE POLICIES AND THEIR COMPARISON

In a Public Key Infrastructure, a certificate authority issues an end user, an X.509 version 3 certificate according to one or more given certificate policies. Assume that there are two differ-ent companies, each with its own PKI and they like to merge and unify their infrastruc-tures[20][21]. For that, they have to establish a connection between the two different domains. Cross certification is a possible solution. If the certificate policies are different,policy mapping is a solution to establish secure communication through unified domains. Issues regarding un-ification of the policies is mentioned in RFC 3647[5]. This paper describes the problems and the solution approaches for it. We explain a procedure for comparison and assessment of Certi-ficatePolicies[22]. The methodis to calculate a compatibility score, which is the base to decide if the unification of the PKIs is possible or not. If unification is possible, a prototype of a uni-fied certificate policy for the merging companies can be created. The CPs must be standardized before they can be compared. Throughout the paper, wherever the term "certificate policy" is mentioned, it is assumed that the certificate policy is the standardized one. The standardized certificate policy file is in the TXT format that can be the input for a simple parsing process. The content of the Certificate Policy is divided into paragraphs, subparagraphs, options, con-nectives etc.

First of all, the outline of the certificate policies will be parsed. Thecomparison of these CP out-lines of different PKIs is possible by parsing. The parsing is carriedout according to some syn-tactical rules. Parsing the outline of certificate policies isproposed in RFC 3647[5]. Two CPs are parsed and compared. Based on the degree ofcomparisona compatibilityscore is calculated which is described in section V. The Figure 8shows the flowgraph of the parsing process.

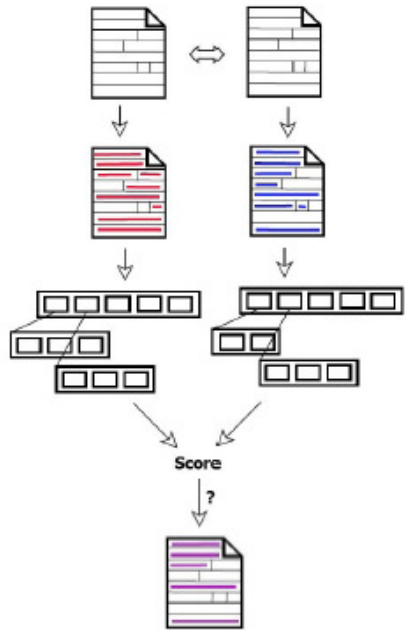

Figure 8. Flowgraph of the parsing process

As shown in Figure 8, the CPs of two PKIs are taken as the input for the parsing process and different tokens from each CP are stored in a tree data structure for comparison. Based on the compatibility score, a prototype of a unified certificate policy for the merging PKIs will be created.

## 4.1 Parsing Certificate Policies

A certificate policy is prepared in the TXT format. The following lines represent a part of a typical certificate policy.

```
1   INTRODUCTION
  1.1 Overview
       //Gives an overview about the document
  1.2 Document name and identification
      a ) RECOMMENDED Document name
      b) MUST Designated identification
     Connection AND
  1.3 PKI participants
     //Described in the subsections
    1.3.1 Certification authorities
         a ) Issues certificates to end users
         b) Issues certificates to other users
         1.3.1.1 Root authorities
                a ) Specifies the difference  to the CAs
    1.3.2 Registration authorities
    1.3.3 Subscribers
    1.3.4 . . .
```

In order to parse a certificate policy, it is represented in Backus Naur Form(BNF) or Extended Backus Naur Form(EBNF)[29]. The syntax of the certificate policy is described in Extended Backus Naur Formnotation in the followingway:

> CertificatePolicy = {Section}
> Section = Point{MainSection | SubSection}
> Point = Number " . "
> MainSection = {CapitalLetter ["("Number")"]} [Weight ] "\n" Content
> SubSection = {Point} Number String [Weight ] "\n" Content
> Number = Integer
> Weight = Integer
> Content = {Option | [ Connection ]}|{Extension}
> Option = [letter ")" ] [ Keyword ] String "\n"
> Connection = "connection" ("AND" |"OR")
> Extension = Subsection
> Keyword = "MUST" | "RECOMMENDED" | "OPTIONAL" | "NOT"

A symbol can either be a number, a dot or a string. Number means the enumerationof the current paragraph or the subparagraph, and the string stands for the currenttitle of the paragraph. The dot is used as the separator in an enumeration. Theparser will check both the title and content of a paragraph. After parsing the certificate policies, the tokens arestored in designated data structures. Two data structures "Token" and "TokenList" are introduced. "Token" represents a simple entry, "TokenList" contains all entries in the same level under a section,e.g. the main token list, which contains all main sections. At the end the content ofa certificate policy is stored as a tree data structure as shown in Figure 9.

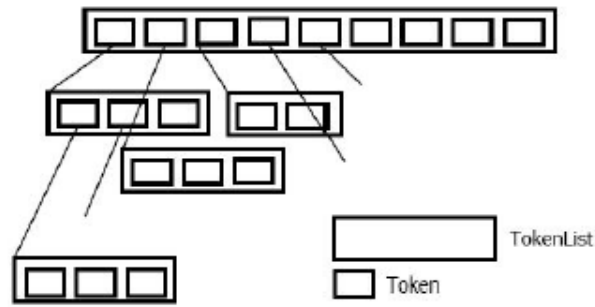

Figure 9. Tree structure representing a certificate policy

## 5. FORMALIZATION OF CP COMPARISON AND ASSESSMENT

The CP comparison and assessment tool is developed in Java with OpenSSL and NetBeans 6.7.1 as the IDE. In the implementation, a token represents a paragraph and a token list represents the list of paragraphs. Under every paragraph there are subparagraphs and up to four levels of indexing is considered. A Token list is a collection of tokens. Let $P_A$ be an arbitrary paragraph of a standardized certificate policy of an entity A and $P_B$ be the corresponding paragraph of the policy of entity B. $S_i$ is the score of a paragraph(token) i in a list. $l_i$ is the score of the subparagraphs(list) under the paragraph i, with i = 1, ...,N where N is the number of paragraphs. $O_{jk}$refers to the provisional result of the equality of option j of $P_A$ with option k of $P_B$. $v_j$ refers to the keyword value of option j in $P_A$, and $v_k$is the keyword value of option k in $P_B$. Each option in the paragraph of policy A is compared with the same paragraph of B's policy. Basically, the value of $O_{jk}$is 100 if the corresponding options in both the policies are same and 0 if they are different. In general,

$$O_{jk} = \begin{cases} 100 \ if \ option \ j \ of \ P_A \ equals \ option \ k \ of \ P_B \\ 0 \ otherwise \end{cases}$$

## 5.1 Weight of a paragraph

The updated weight of a paragraph idepends upon the size of the TokenList. If there are many elements in the paragraph, i.e., if the size of the TokenList is large, the updated token score gets lessweight as shown in equation (1).

$$S_i = (S_i + l_i*N)/(I+N) \qquad\qquad (1)$$

For example,
If the original paragraph score, $S_i$, is 100,subparagraph score is 75 and the number of elements(subparagraphs) under that paragraph is 2, then the updated score is

$S_i$=(100+75*2)/(1+2)=83.3

If the number of elements is increased to 8, then

$S_i = (100+75*8)/(1+8)=77.8$

Thus, as the number of elements increase, the paragraph weight decreases.

## 5.2 Merging PKIs based on policies

The calculation of the score of a token depends upon the following two situations:
- *$P_A$ has some options, $P_B$ has no options:*

    In this case, merging A and B with cross certification is not allowedsince the score $S_i$ is0.

However, if A acquires B, B can take the options ofA and the score $S_i$is 100.

- *Both $P_A$ and $P_B$ have some options:*

    In this case, for merging A and B with cross certification, $S_i$ is calculatedas the ratio of sum of the partial results of the options to themaximum number of options from both sides as in equation (2).

$$S_i = \frac{\sum O_{jk}}{Max(Numberof Options(P_A), \ Numberof Options(P_B))} \qquad (2)$$

If A acquires B, the sum is divided by the number of options of $P_A$ as shown in in equation (3).

$$S_i = \frac{\sum O_{jk}}{Numberof Options(P_A)} \qquad\qquad (3)$$

## 5.3 Overall CPs compatibility Score

Let us consider the certificate policies $CP_A$ and $CP_B$ that belong to the organizations A and B respectively.Let us also consider that A is the comparing organization, and, therefore, it is the onethat sets the criteria for the CPs comparison (or else the one that decides aboutpossible compatibility with B). This means that, the CPs comparison is actually theassessment of $CP_B$ against $CP_A$ using the specific criteria of A.

The overall compatibility score(total score) of $CP_B$ against $CP_A$ is extracted as aweighted average of all CP paragraphs scorings, as in equation (4):

$$C = \frac{\sum (S_i * W_i)}{\sum W_i} \qquad (4)$$

where C defines the overall compatibility score of $CP_B$ against $CP_A$, $W_i$ defines theweight of CP paragraph $P_i$ and $S_i$ defines the scoring for CP paragraph $P_i$. Theaddition $\sum(S_i * W_i)$ is the weighted compatibility score for the two CPs, whereas$\sum W_i$ is the maximum possible overall compatibility score of the two CPs.
The total score can also be calculated using equation (5):

$$C = \frac{\sum S_i}{N} \qquad (5)$$

where N is the number of paragraphs.

## 5.4 Keywords

Some keywords are also helpful to prepare the Certificate Policy. The proposed key words introduced in RFC 2119[30] are MUST, MUST NOT, REQUIRE, SHALL, SHALL NOT, RECOMMENDED and MAY / OPTIONAL.

Only few of the proposed keywords are used in the implementation. They are: MUST, RECOMMENDED, OPTIONAL, and NOT.

The options of a paragraph can look like: *KEYWORD phrase*

Basically the keywords are interpreted by numerical values between 0 and 1.

|              |     |
|--------------|-----|
| MUST         | 1.0 |
| RECOMMENDED  | 0.8 |
| OPTIONAL     | 0.5 |
| NOT          | 0.0 |

Since these values are fuzzy in nature, we can get a token score from Fuzzy Logic[31].The idea is that, the score should not be 0 or 100 strictly, it shall be possible to getscores in between 0 and 100. Fuzzy logic gives a degree of truth, which is between 0 and 1. Toadopt this idea, the degree of similarity of the keywords will be taken.The keyword values have certain meanings. "MUST" indicates an absolute necessaryoption and has the value 1.0. On the other side, "NOT" with 0.0 indicates a non-desired option. "OPTIONAL" means, the option may or may not occur and has the value 0.5. "RECOMMENDED" is closer to "MUST" than to "OPTIONAL", so the value is 0.8. Thekeywords and their values provide an exact determination of equality and a betterexpression of the need of a certain option. They also state a kind of degree of truth. Ifthe same option is found at both paragraphs $P_A$ and$P_B$, and the keywords are similar,like "MUST" and "RECOMMENDED", then the paragraph score is high. Therefore the difference in the keywords will be subtracted by 1 and the result is multiplied with100 (score of equality).

## 5.5 Connectives with Options

Connectives like, OR and AND can be used with options. With an OR connective, we choose one option out of many. It is mathematically represented as the maximum value of the provisional scores of the equal options as in equation (6).

$$S_i = Max(O_{jk} * (1 - |v_j - v_k|)) \qquad (6)$$

With an AND connective, all the options have to be chosen. So it has to be determined that how many options are equal with what degree. The score can be calculated as in equation (7)

$$S_i = \frac{\sum O_{jk} * (1 - |v_j - v_k|)}{Max(Number of Options(P_A), \ Number of Options(P_B))} \qquad (7)$$

The equation (7) refers to the policies that should be merged with cross certification. If A acquires B, the denominator is the number of options of $P_A$, because B has to adapt to A. Therefore the number of options of B are irrelevant as in equation (8).

$$S_i = \frac{\sum O_{jk} * (1 - |v_j - v_k|)}{Number of Options(P_A)} \qquad (8)$$

For example, consider the options in the paragraphs $P_A$ and $P_B$as shown below:

| $\underline{P_A}$ | $\underline{P_B}$ |
|---|---|
| a) MUST a | a) RECOMMENDED a |
| b) MUST b | b) OPTIONAL b |
| c) MUST c | c) RECOMMENDED d |
| | d) RECOMMENDED e |

Suppose the connection is OR, then both for merging as well as acquisition of companies,

$S_i$=Max{100*(1-(1.0-0.8)), 100*(1-(1.0-0.5)), …, 0} = 80

Suppose the connection is AND and the requirement is that A and B are to be merged with cross certification, then

$S_i$= 80/4+50/4+0+0=32.5

Suppose the connection is AND and A acquires B, then
$S_i$ =80/3+50/3+0=43.3

## 5.6 Final acceptance rules

The final acceptance rules are set by the comparing organization and are those thatdetermine whether the foreign CP can be accepted or rejected, after the CPs comparison has been performed. Examples of such rules, could be a combination of statementslike: "the overall CP compatibility score should be over 90%", "partial compatibilityin paragraph X should be over 80%" etc. It is also possible to set as only final acceptance rule that the partial CP compatibility in one (or more) paragraph is 100%,which means that a foreign CP may be rejected solely because of incompatibility in this specific paragraph. Based on this information, the final acceptance rules canreflect any acceptance or rejection policy that the comparing organization wants tofollow and, thus, make the method very flexible to different needs and requirements.

## 6. CONCLUSION

The certificate policy indicates the PKI certificates' profile and the architectural structure of the underlying trusted third party. Thus a certificate authority issues a certificate to an end user upon a specific certificate policy. If different end users with different domains want to establish a secure communication, they need to find a wayhow they can trust each other. That means their certificates need to follow the samestandard. This paper presents a novel method for comparison and assessment of Certificate Policies for their unification during merger and acquisition of companies. The final acceptance rules are set by the comparing organization and are those that determine whether the foreign CP can be accepted or rejected after the CPs comparison has been performed. Unification of CPs is allowed only if the compatibility score satisfies the final acceptance rule.

## REFERENCES

[1] A. Arsenault and S. Turner. (2002) Internet X.509 public key infrastructure: Roadmap, internet draft. PKIX Working Group, IETF. [Online].Available: http://ieft.org/internet-drafts/drafts-ieft-pkix-roadmap-08.txt

[2] R. Housley,W. Ford,W. Polk, and D. Solo. (1999) Internet X.509 public key infrastructure certificate and CRL profile, IEFT RFC standard 2459.[Online]. Available: http:www.ieft.org/rfc/rfc2459.txt

[3] *Information Technology—Open Systems Interconnection—The Directory:Authentication Framework*, CCITT Rec. X.509/ISO/IEC Standard9594-8, 1994.

[4] G. A. Weaver, S. Rea, and S. W. Smith, "A computational framework for certificate policy operations", Dartmouth Computer Science Technical Report TR2009-650, Dartmouth College, Jun 2009.

[5] S. Chokhani, W. Ford, R. Sabett, C. Merrill, and S. Wu, "RFC 3647: Internet X.509 public key infrastructure certificate policy and certification practicesframework", Nov 2003.

[6] S. Chokhani and W. Ford, "RFC 2527: Internet X.509 public key infrastructure certificate policy and certification practices framework", Mar 1999.

[7] A. Bourka, D. Polemi, and D. Koutsouris, "Interoperability among healthcare organizations acting as certificate authorities", IEEE Transactions on InformationTechnology in Biomedicine, vol. 7, pp. 364-377,Dec 2003.

[8] Cross Certification Guidelines, Canadian institute for health informatics (CIHI),CA. [Online]. Available: http://secure.cihi.ca, 2001.

[9] Cross certification methodology and criteria, tech. document, Treasury BoardSecretariat of Canada, Government of Canada, Ottawa, Canada, 2001.

[10] Z. Kardasiadou, B. Blobel, and S. Amberla., "Legal and policy issues of pkiadoption in health telematics applications in Greece, Germany and Finland",Tech. Del. D.2.2. RESHEN Project, European Commission. [Online]. Available:http://www.biomed.ntua.gr/reshen, 2001.

[11] Policy requirements for certification authorities issuing qualified certificates,ETSI TS 101 456. [Online].,2000, Available: http://portal.etsi.org/sec/el-sign.asp

[12] Policy requirements for certification authorities issuingqualified certificates, ETSI TS 101 456, 2000. [Online]. Available:http://portal.etsi.org/sec/el-sign.asp

[13] K. Louwerse, F. A. Allaert, B. Blobel, and B. Barber, *Security Standardsfor Healthcare*. Amsterdam, The Netherlands: IOS Press, 2002.

[14] R. Fraser, "Healthcare PKI standards development," in *2001 CanadianInstitute for Health Information Workshop*, Ottawa, Canada.

[15] *Healthcare Informatics—Public Key Infrastructure*, ISO TS 17090 Parts1-3, 2001.

[16] Healthcare certificate policy, ASTM standard E31.20 (2000). [Online].Available: http://www.cio.gov/fpkisc

[17] Directive 1999/93/EC of the european parliament and of the council of 13 December 1999 on a community framework for electronic signatures, Official J., vol. L 013, pp. 00120020, Jan. 2000.

[18] Heng Pan et al., An Efficient Scheme of Merging Multiple Public Key Infrastructures in ERP, LNCS, Springer-Verlag Berlin, 2005, pp. 919-924

[19] Lloyd et al., CA-CA Interoperability, PKI Forum, 2001. http://www.pkiforum.org/pdfs/ca-ca interop.pdfof Business Administration Asahi University, Japan.

[20] Peter M. Hesse, David P. Lemire, "Managing interoperability in non-hierarchical public key infrastructures" Gemini Security Solutions.

[21] Zheng Guo, Tohru Okuyama, Marion R. Finley. Jr., "A New trust Model for PKI Interoperability", Information Network Laboratory, Graduate School.

[22] Balachandra, Prema K.V., "Design of a Public Key Infrastructure to handle Interoperability Issues" , a Ph.D thesis, Manipal University, Manipal, India, December, 2012.

[23]Satoshi Koga, Kouichi Sakurai, "A Merging Method of Certification Authorities Without Using Cross-Certifications", Proceedings of the International Conference on Advanced Information Networking and Application (AINA'04) 2004 IEEE

[24] M. Cooper et. al., Internet X.509 Public Key Infrastructure : Certification Path Building, Network working group, RFC 4158, September, 2005

[25] John Linn, Trust Models and Management in Public-Key Infrastructures, RSA Laboratories, 2000

[26] Cristina Satizbal, Juan Hernndez-Serranoa,Jordi Forna, and Josep Peguerolesa, "Building a virtual hierarchy to simplify certification path discovery in mobile ad-hoc networks, Computer Communications", Volume 30, Issue 7, 26 May 2007, Pages 1498-1512

[27] Cristina Satizbal, Rafael Pez, Jordi Forn, "Building a Virtual Hierarchy for Managing Trust Relationships in a Hybrid Architecture", Journal of Computers, VOL. 1, NO. 7, October/November 2006.

[28] S. Adams, and S. Farrell, "Internet X.509 Public Key Infrastructure Certificate Management Protocols, Network Working Group Request for Comments 2510", 1999. http://www.ietf.org/rfc/rfc2510.txt

[29] N. Wirth, Compiler Construction, ch. 2. Nov 2005.

[30] S. Bradner, RFC 2119: Key words for use in RFCs to indicate requirement levels, Mar 1997.

[31] G. J. Klir and B. Yuan, Fuzzy sets and fuzzy logic: theory and applications. Prentice Hall, 1995.